\documentclass[letterpaper] {article}     
\usepackage{cite}           
\usepackage{graphics}
\usepackage{ifthen}
\usepackage{epsfig}
\usepackage{amsmath}
\usepackage{amssymb}
\usepackage{color}

\begin{document}

\bibliographystyle{unsrt} 

\title{Double-Spin Asymmetry in the Cross Section for \\
Exclusive   $\rho^0$  Production in  Lepton-Proton Scattering  }
\maketitle              %$
\begin{center}          %$
%* \author
{ { \it The HERMES Collaboration }\medskip \\
A.~Airapetian,$^{31}$
N.~Akopov,$^{31}$
Z.~Akopov,$^{31}$
M.~Amarian,$^{23,26,31}$
J.~Arrington,$^{2}$
E.C.~Aschenauer,$^{7}$
H.~Avakian,$^{11}$
R.~Avakian,$^{31}$
A.~Avetissian,$^{31}$
E.~Avetissian,$^{31}$
% ...
}

\pagebreak

%% \institute{ 
% ...
$^2$Physics Division, Argonne National Laboratory, Argonne, Illinois 60439-4843, USA\\
% ...
$^7$DESY Zeuthen, 15738 Zeuthen, Germany\\
% ...
$^{11}$Istituto Nazionale di Fisica Nucleare, Laboratori Nazionali di Frascati, 00044 Frascati, Italy\\
% ...
$^{23}$Nationaal Instituut voor Kernfysica en Hoge-Energiefysica (NIKHEF), 1009 DB Amsterdam, The Netherlands\\
% ...
$^{26}$Istituto Nazionale di Fisica Nucleare, Sezione Roma1-Gruppo Sanit\`{a} and Physics Laboratory, Istituto Superiore di Sanit\`a, 00161 Roma, Italy\\
% ...
$^{31}$Yerevan Physics Institute, 375036, Yerevan, Armenia\\
%% } 

%\date{Received: \today / Revised version:}
%\titlerunning{Paper Tag: RHO2S}
%\authorrunning{The HERMES Collaboration}
\end{center}

\begin{abstract}        
  Evidence for a  positive longitudinal double-spin asymmetry
  $  \langle A_1^\rho \rangle = 0.24 \pm 0.11_{\rm stat} \pm 0.02_{\rm syst} $
  in the cross section for   exclusive diffractive  $\rho^0(770)$   vector meson production 
  in polarised    lepton-proton scattering 
  was observed by the  HERMES experiment. 
  The longitudinally  polarised 27.56~GeV   HERA positron beam  was scattered  off a 
  longitudinally polarised pure hydrogen gas target. 
  The   average    invariant mass of the photon-proton system has a value of 
  $\langle  W \rangle = 4.9$~GeV,  while the  average negative  squared four-momentum
  of the virtual photon  is   $\langle Q^2 \rangle = 1.7$~GeV$^2$.
   The ratio of the present result to the corresponding spin asymmetry in inclusive 
  deep-inelastic scattering   is in agreement with an early  theoretical 
   prediction based on the generalised vector meson dominance model. \\
PACS numbers : 13.25.-k; 13.40.-f; 13.60.-r; 13.60.Le; 13.88.+e;  14.40.Cs \\
Keywords : Lepton-Nucleon Scattering, Rho Production, Asymmetries, Photoabsorbtion. 
\end{abstract}

\twocolumn

\begin{figure}[!ht]
  \includegraphics[width=7.5cm,height=2.7cm]{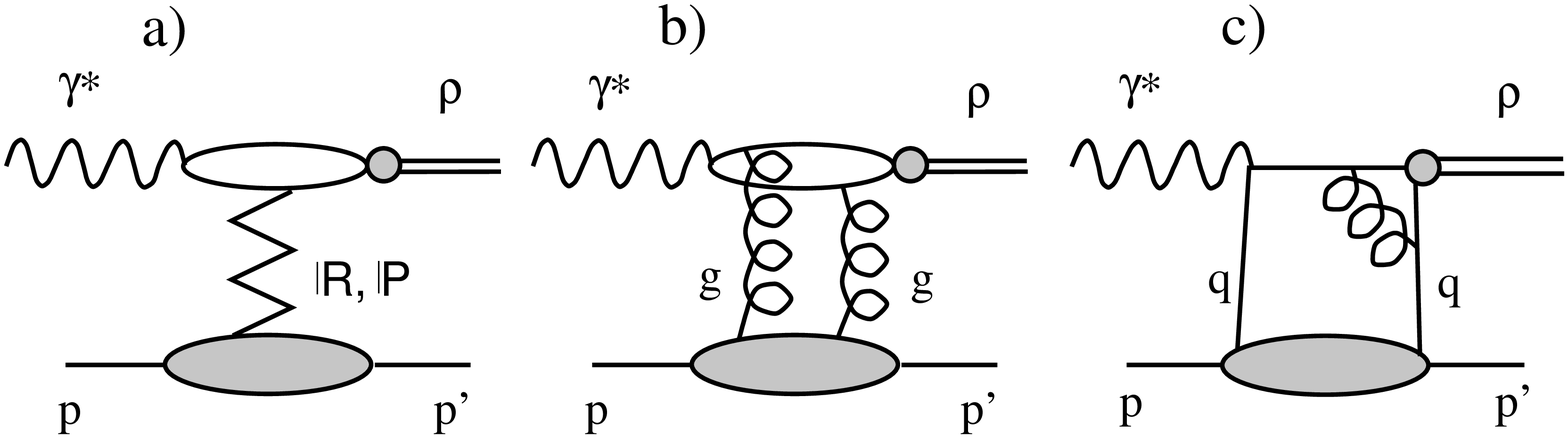}        
  \caption{Schematic graphs for various models of exclusive 
    diffractive $\rho^0$  production: a) Reggeon or Pomeron exchange 
    in models based on Regge theory,  b) two-gluon  exchange and c)
        quark exchange  
    in models inspired by perturbative QCD.
    \label{fig:vmd_models}}
\end{figure}

Diffractive $\rho^0$ production 
in lepton-nucleon \linebreak scattering  
is often  described as the fluctuation of the  virtual photon emitted by the lepton
into an intermediate virtual $q \bar{q}$ state
or off-shell $\rho^0$  meson.
This intermediate state is 
scattered onto the
mass shell by a  diffractive strong 
interaction with the target,
leaving the target nucleon intact~\cite{Sakurai}.
Several competing models for this process are
schematically shown  in the graphs of Figure \ref{fig:vmd_models}.
At low energy  the $\rho^0$ cross section shows a  strong decrease with 
increasing energy~\cite{Bauer:1978,Cassel:1981,Haakman:1996},
which  can be described  by  the  exchange of Reggeons.
At an invariant mass of the  photon-nucleon system of approximately
$W\!=\!5$~GeV   (see below for definitions of all kinematic variables), 
the  cross section exhibits a dramatic change from a rapidly falling to a  weakly rising  
$W$-dependence~\cite{Crittenden:1997,Hermes:00rhoxsect}.
Above this energy, models of the
interaction based on Regge theory involve the 
exchange of Pomerons.
 Alternatively, there exist
 perturbative QCD calculations of vector-meson production by 
longitudinal photons that are based on the exchange of quarks 
and gluons and on the non-perturbative description of 
 nucleon structure in terms of skewed parton 
distributions~\cite{rady:1996,XJi:1997,Collins:1997,vanderh}.
Both  types of models 
--- Regge theory and pQCD calculations ---
have achieved some  degree of success  in reproducing  the observed unpolarised
cross sections for $\rho^0$  production~\cite{Hermes:00rhoxsect,Zeus:rho,H1:rho}. 
It has been shown that in  the HERMES kinematic range,  the data 
for exclusive $\rho^0$ production by longitudinal photons can 
be described by a  pQCD calculation
that includes  a combination of  both quark and gluon 
 exchange mechanisms~\cite{Hermes:00rhoxsect}.

Previously,
the spin dependence of the $\rho^0$  leptoproduction process 
has been investigated by measuring the angular distributions of the
production and the 
self-analysing $\rho^0 \rightarrow \pi^+ \pi^-$ decay.
Spin degrees of freedom in the cross section 
can be  described by   spin density matrix elements constructed from 
helicity-conser\-ving and non-conser\-ving amplitudes for
particle exchange in the $t$-channel~\cite{wolf:1973}. 
Experiments have shown that the helicity of the 
photon in the $\gamma^* N$ centre-of-mass system
is approximately retained by the $\rho^0$ meson, a phenomenon
known as $s$-channel helicity conservation (SCHC),
and that the exchanged object has natural parity $(-1)^L$~\cite{Bauer:1978},
which can be associated with {\it e.g.} 
Reggeon  and Pomeron exchange.
Typically,  the initial spin states  of the target nucleon 
were  averaged and  the final spin states were summed~\cite{wolf:1973}, 
since they were experimentally inaccessible.
The general  case where  the  initial spin states  of  a  longitudinally polarised beam 
and a longitudinally or transversely  polarised  target   are   explicitly  included 
in the formalism of  spin density matrix elements  has been discussed in Ref.~\cite{Fraas:1974}.

Up to now, little attention has been paid to the 
theoretical prediction of double-spin asymmetries in  the cross section for 
diffractive processes, i.e. to the dependence of the cross section on the product of  the initial 
polarisations of beam {\it and} target; usually a double-spin  asymmetry of zero was assumed for 
$\rho^0$ production.
Nevertheless, there exists  early work where 
spin asymmetries are  given by the ratio of helicity-conserving 
amplitudes of unnatural to  natural parity exchange in the $t$-channel;
this work is based on the formalism of spin 
density matrix elements  and  the generalised vector meson dominance model (GVMD)~\cite{Fraas:1975gvmd}
which describes  the hadronic fluctuation of the virtual photon as a coherent  superposition of  vector meson states and the transitions between them.
In particular,  the  longitudinal  double-spin asymmetry in exclusive $\rho^0$ production  was predicted to be about twice as large as the corresponding 
asymmetry in inclusive deep-inelastic scattering (DIS)~\cite{Fraas:1976asym}.

No published  prediction based on  perturbative QCD calculations
exists for double-spin asymmetries in the photo- or lepto-production of $\rho^0$ mesons.
Polarised quasi-real  $J/\psi$ photoproduction 
involving  the exchange of two gluons  was discussed 
in Ref.~\cite{vaenttinen:1998}.
However,  because this calculation relies on the heavy quark approximation, 
the relevant physics for this process is qualitatively  different
and  no direct implications for exclusive $\rho^0$
production  in the HERMES kinematic range can be drawn.

This paper 
presents the first observation 
of a non-zero longitudinal  double-spin  
asymmetry in the cross section for  exclusive   $\rho^0(770)$ meson production in 
polarised lepton-nucleon scattering. \linebreak
This  observable was measured by the HERMES experiment~\cite{Hermes:98spec}
in the years 1996 and 1997. \\

The HERMES  experiment  uses  the  polarised 
27.56~GeV  positron beam of the HERA storage ring. 
A transverse polarisation of the posi\-tron beam develops
through an asymmetry in  the small spin-flip amplitudes for
synchrotron radiation in the bending dipoles --- 
the Sokolov-Ternov effect\linebreak~\cite{Sokolov-1964}.
Longitudinal beam  polarisation is  achieved  by   spin rotators 
in front and behind  the experiment.
Typical beam  polarisation values are between $0.5$ to $0.6$, measured with a
negligible statistical uncertainty and a 
systematic uncertainty of 0.02~\cite{bar}.

In the years 1996 and 1997 a longitudinally polarised internal atomic hydrogen 
gas target was used~\cite{ste}.
The orientation of the  target polarisation is randomly 
selected  about once per minute. 
The average target polarisation 
for the  combined 1996 and 1997 data  on  the  polarised hydrogen target 
is $0.88 \pm 0.05$,
where the uncertainty is predominantly systematic.

The   forward  magnetic spectrometer~\cite{Hermes:98spec} is
divided into  symmetric upper and lower 
halves  by the positron and the (unused) proton beam lines.
The acceptance covers 
$40 \!< \!|\theta_v |\!< \! 140$~mrad in the vertical direction and
$|\theta_h|\! < \! 170$~mrad in the horizontal direction.
Over the kinematic range of the experiment, the momentum 
resolution  for positrons  is 0.7-1.3\,$\%$ and the  uncertainty in the
scattering angle is about 0.6\,mrad. 
Positrons are distinguished from hadrons by  four sub-systems 
for particle identification:
a lead-glass electromagnetic  calorimeter, a preshower detector, 
a tran\-sition-ra\-di\-a\-tion detector, and 
a   threshold \v{C}eren\-kov detector.
The positron identification
has an efficiency of better than 98$\%$ at
a hadron  contamination of less than $1\%$ over the
kinematic range of the experiment.
The  relative luminosity for the two target 
spin states is measured  by counting coincident $e^+e^-$ pairs 
from elastic (Bhabha)  scattering  of the beam positrons
off  the  electrons of the target gas atoms.
The present data correspond to an integrated luminosity 
of about 55~pb$^{-1}$. \\

At HERMES energies, the  lepton-nucleon interaction
is mediated  by  a  virtual photon with negative 4-momentum squared 
$Q^2 \! \equiv \! -q^2\! \equiv \! -\!(k\!-\!k')^2  \!\sim\! 4EE'\sin^2\left({\theta}/{2}\right)$,
where $k\,(E)$ and $k'\,(E')$ denote  the four-momenta (energies) of the 
incoming and outgoing lepton 
and $\theta$ is the polar scattering angle of the lepton.
The target nucleons are at  rest: $p=(M,\vec{0})$.
The  invariant mass $W$ 
of the photon-nucleon system is given by   
$W^2 \! \equiv  \!(q+p)^2 \! \stackrel{lab}{=}\!  M^2 \!+\! 2M\nu \!-\!Q^2$.
Here, $\nu \!\equiv\! {p\cdot q}/{M} \!=\!E\!-\!E' $ denotes the photon energy    in the target rest frame 
and $y  \equiv  \nu/E$ the fractional photon energy.
The Bj{\o}rken scaling variable is defined as  $x\! =\! Q^2/2M\nu$.
In diffractive $\rho^0$ production, the photon-nucleon 
cross section falls exponentially with 
the squared four-momentum  transfer to the target
\mbox{$t \!  \equiv\! (q\!-\!v)^2 \!<\! 0$}, 
with $v$ the four-\-mo\-men\-tum of the $\rho^0$ meson.
At $t_0$, the maximum (least negative) value of $t$  kinematically allowed for 
fixed  $Q^2$, $\nu$, $M_Y$,  and $M_{\pi\pi}$, the momentum of the final 
state $\rho^0$  meson  is parallel to the direction of the incoming photon in 
the photon-nucleon  centre-of-mass system.
Here,  $M_Y$  is the invariant  mass of the undetected  final state and 
$M_{\pi\pi}=\sqrt{v^2}$
 is the reconstructed invariant  mass of the  $\rho^0$ candidate.
The   squared four-momentum transfer $t'$ beyond   $t_0$   is given by
$t' \!\equiv \!t\!-\!t_0 \!< \!0$. 
In order to select  exclusive diffractive
events,  the   excitation energy
$\Delta E$ transferred to the target nucleon  
can be used as a measure of exclusivity: 
$ \Delta E \!\equiv \!{(M^2_Y \!-\! M^2)}/{2M} \!\stackrel{lab}{=}\!\nu\!-\!E_{\rho}\! + \!{t}/{2M}$, with 
$E_{\rho}$ the energy of  the  $\rho^0$  meson. In the case 
of an exclusive   process,  no  energy  is transferred to the 
target ($\Delta E \cong 0$),   and  the target nucleon stays
intact (${p'}^2=p^2$).

Three angles $\Phi$, $\phi$, and $\theta$  are necessary  for a 
complete description of  the angular distribution 
of the $\rho^0$ meson production and decay~\cite{wolf:1973}.
The azimuthal production angle $\Phi$ is the angle between the lepton
scattering plane and the $\rho^0$  meson production plane in the
photon-nucleon  centre-of-mass frame.
The  $\rho^0$  decay is described in the $\rho^0$ meson rest frame 
by two angles: i) the azimuthal angle $\phi$ between the production 
and  the decay plane and ii) the  polar angle $\theta$ of the
positively charged decay particle with respect to the $z$-axis 
of the $\rho^0$ meson rest frame, which is defined  opposite to
the direction of the  scattered nucleon. \\

In  the exclusive process $\mathrm{ep} \rightarrow \mathrm{ep} \rho^0$,
only the  scattered positron and the   $\rho^0\rightarrow \pi^+\pi^-$  decay products
are detected at HERMES,  since the recoiling target proton remains outside of
the spectrometer acceptance. 
Consequently, as a first step,  only events  were selected with exactly 
three tracks --- 
a positron and  two
oppositely charged hadrons.
The tracks are required to be  within the nominal spectrometer
acceptance and  to originate from a common vertex in the  target region.
To ensure  that the  trigger  efficiency is  close to unity,
a minimum energy of the scattered positron
above the  calorimeter threshold of the trigger  
is required. 

\begin{figure}[!th] 
  \includegraphics[bbllx=-20pt,bblly=25pt,bburx=342pt,bbury=342pt,width=6.cm,height=4.cm]{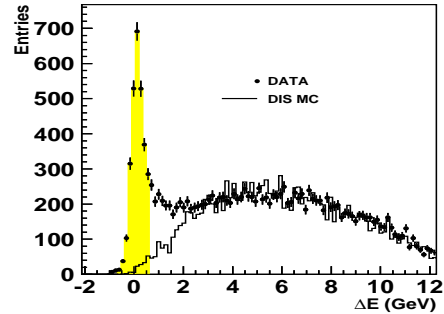}
  \caption{  \label{fig:rho_de} 
    Distribution in the excitation energy $\Delta E$
    for the decay channel \mbox{$\rho^0 \rightarrow \pi^+\pi^-$}, 
    after all other event selection criteria were applied.
    The histogram is a Monte Carlo simulation 
    of combinatorial background from deep-inelastic scattering (DIS).
    The shaded area indicates the events that were used for the analysis.}
 \end{figure}

The $\rho^0$ candidates are selected within the range  \linebreak
\mbox{$0.62 \!<\! M_{\pi\pi} \!< \! 0.92$~GeV} of
the  reconstructed  
invariant mass of the  hadron pair.
The  requirement 
$-t' < 0.4 $~GeV$^2$
suppresses non-diffractive processes, which fall off more slowly 
with \mbox{$-t'$}  than diffractive $\rho^0$ production.
Exclusive events appear as a  peak near zero  
in the excitation energy  distribution
shown in Fig.~\ref{fig:rho_de}; 
the shaded area
$| \Delta E| < 0.6$~GeV
indicates  the 
events used for analysis.  For 
\mbox{$\Delta E \gtrsim 2$~GeV} the spectrum is dominated 
by background  from non-exclusive  processes where, for example,  
energy is absorbed by   the target.
After all selection criteria were applied,  about 2800 exclusive  $\rho^0$ events 
remained  from  the  combined 1996 and 1997 data on the polarised 
hydrogen target.  
The average values of the relevant  kinematic variables are
$\langle W  \rangle  \!=\! 4.9 {\rm \; GeV}$,
$\langle Q^2  \rangle  \!=\! 1.7 {\rm \; GeV}^2$,
$\langle  -t' \rangle  \!=\! 0.1 {\rm \; GeV}^2$,
$\langle x    \rangle  \!=\! 0.07$.
HERMES results 
on the  cross section for exclusive $\rho^0$ production and spin density matrix elements 
as well as  distributions of the invariant mass, $t'$ and other  kinematic variables  can be found 
in Refs.~\cite{Hermes:99coherence,Hermes:00rhoxsect,Hermes:99angles}.

The  predominant  contribution  to the back-\linebreak ground from 
non-exclusive processes  is combinatorial hadronic 
background from   deep-inelastic scattering events.
In the region of the exclusive  peak  (shaded area in Fig.~\ref{fig:rho_de}),  it  is  not separable  on an event-by-event basis.
It is  subtracted using a Monte Carlo simulation based 
on the LEPTO~\cite{LEPTO} generator and the LUND fragmentation model~\cite{LUND:94}.
The  Monte Carlo events are  subject to the $\rho^0$ event selection 
criteria yielding the $\Delta E$ distribution  shown by the histogram 
in Fig.~\ref{fig:rho_de};
it is normalised to the data   
in the region  \mbox{$\Delta E > 3$~GeV}. 
The  contamination in the $\rho^0$ data sample 
within the signal region $|\Delta E|\! < \!0.6$~GeV
is typically less than $10\%$.
For  the asymmetry analysis, the background was subtracted 
separately for each  spin state and  kinematic bin, accounting
for  the asymmetry contribution from DIS   background.
The  statistical uncertainty in this  background correction 
is propagated into the statistical uncertainty of the 
measured asymmetry.

An additional contribution to the  non-exclusive background arises  
from double-dis\-sociative   diffractive  $\rho^0$ production.
This process is  similar to the exclusive process of interest,
except that the proton target is  dissociated.
Based on  previous  measurements~\cite{E665:1997},
the contribution  of this background  
was estimated~\cite{Hermes:00rhoxsect} to be less than 
$6 \pm 2$\,\%  within the stringent  $|\Delta E| < 0.6$~GeV
requirement.

Background from exclusive processes includes 
the contributions of non-resonant pion pair production
and of the  decay  $\omega(783)\rightarrow \pi^+\pi^-$ 
(branching ratio  $2.2\%$). 
These two processes contribute less than $1\%$
to the signal region; for the present  analysis both contributions remain within 
the data sample.
Mis-reconstructed $\phi$ meson decays 
$\phi \rightarrow \mathrm{K}^+\mathrm{K}^-$
appear at $M_{\pi\pi}$  below 0.6~GeV, and are 
excluded by the invariant mass requirement. \\

\newcommand{\lpar}{{\stackrel{\rightarrow}{\scriptscriptstyle \rightarrow}}}
\newcommand{\lant}{{\stackrel{\rightarrow}{\scriptscriptstyle \leftarrow}}}
\newcommand{\tpar}{{\stackrel{\rightarrow}{\uparrow}}}
\newcommand{\tant}{{\stackrel{\leftarrow}{\uparrow}}}

In lepton-nucleon scattering, with both  target  and beam  
longitudinally polarised,  the experimentally accessible lepton-nucleon cross section 
asymmetry $A_\|$ is defined in terms of $\sigma^{\lpar}$ and $\sigma^{\lant}$, 
the  cross sections  for  parallel and  anti-parallel orientation of the 
target polarisation  with respect to the direction of the beam polarisation:
\begin{equation} \label{eq:Apardef}
  A_\| \; \equiv \;  \frac{\sigma^{\lant} - \sigma^{\lpar}}
  {\sigma^{\lant} + \sigma^{\lpar}} 
  \; = \;   \frac{ N^{\lant} L^{\lpar} - N^{\lpar} L^{\lant}}
  {N^{\lant} L_{P}^{\lpar} + N^{\lpar} L_{P}^{\lant}}.
\end{equation}
It is  determined from the  number of  events \linebreak
$N^{\lpar(\lant)}$ per beam and target   spin configuration, 
weighted with the    relative  luminosities of  
each spin  configuration $L^{\lpar(\lant)}$ and $L_P^{\lpar(\lant)}$. Here, 
{$L_P^{\lpar(\lant)}\!=\! 
\langle L\!\cdot\! | p_B\! \cdot\! p_T |\rangle^{\lpar(\lant)}$ }
is the relative 
luminosity  weighted with  the product of 
beam and target polarisation.
For the HERMES pure hydrogen gas target, no dilution from  other
material  exists.
The statistical uncertainty  of $A_\|$ is determined by 
the event statistics; the small  statistical uncertainties of 
the luminosity  and the polarisation measurements  are included 
in the systematic uncertainty.

The  spin dependence of the photon-nucleon interaction is characterised by 
two asymmetries   of the  interaction cross section for virtual photons:
i)  the asymmetry $A_1$ 
for  a transverse photon  with well defined helicity interacting with  
a longitudinally polarised target  nucleon  and 
ii)  the asymmetry  $A_2$,  which  arises from the  interference
between transverse and longitudinal photons. Specifically,
\begin{equation}\label{eq:a1-vm}
  A_1 = \frac{\sigma_{1/2}-\sigma_{3/2}}{\sigma_{1/2}+\sigma_{3/2}} \quad 
  {\rm and} \quad A_2 = \frac{\sigma_{LT}}{\sigma_{1/2}+\sigma_{3/2}}.
\end{equation}
Here   $\sigma_{3/2}$ and $\sigma_{1/2}$ denote
the virtual-photon interaction cross sections, with {\small $3/2$} and {\small $1/2$}  
the projection of the total spin of the  photon-nucleon  system  
along  the photon momentum, and 
$\sigma_{LT}\! =\! \sqrt{\sigma_L \!\cdot\! \sigma_T}$ is
the interference cross section   between
 longitudinal and transverse photons.
The  definitions of $A_1$ and $A_2$ are formally independent of 
the  physics process that the virtual photon undergoes.
Note that in inclusive DIS,  $A_1$ and $A_2$ can be  interpreted in terms of  the
polarised structure functions $g_1$ and $g_2$.

The measured asymmetry $A_\|$ is  related  to  the 
photon-nucleon  interaction  asymmetries $A_1$ and $A_2$ by
\begin{equation} \label{eq:apar2a1} 
  A_\|= D \cdot (A_1 + \eta A_2).
\end{equation}
The effective  polarisation $D$ of the virtual photon is given by
\begin{equation} \label{eq:depol_rho}
  D \!=\! \frac{ 1\!-\! (1\!-\!y) \, \epsilon }{ 1 \!+\! \epsilon  R}
  \quad  {\rm and } \quad
  \epsilon \simeq \frac{1-y-\frac{Q^2}{4E^2}}{1 \!-\!y\! +\!\frac{y^2}{2} \!+\!
    \frac{Q^2}{4E^2}},
\end{equation}
with \mbox{$\epsilon \!=\! \Gamma_L/\Gamma_T$}
the    ratio of fluxes and \mbox{$R\! =\! \sigma_L/\sigma_T$} the ratio 
of the reaction cross  sections for  longitudinal and transverse
photons.
The kinematic factor  $\eta$  in Eq.~\ref{eq:apar2a1} is given by
 $\eta \! =\! {2\epsilon \! \sqrt{Q^2}}/\{(M+2E)[1\!-\!(1\!-\!y)\epsilon]\}$.

In exclusive $\rho^0$  production,  the ratio $R$
can be 
measured via  the angular distributions of the
$\rho^0$ decay; it shows a strong 
increase   with $Q^2$, becoming larger than unity \footnote{
For comparison, the ratio $R$ in 
inclusive DIS  has  a flat distribution and varies 
between 0.2 and  0.4~\cite{Withlow:90,Abe:1998}.}
at $Q^2 \!\ge\! 3$~GeV$^2$.
From a fit to HERMES  data~\cite{Hermes:99angles},      
a parameterisation   of $R$ in  the form  
\begin{equation} \label{eq:rrho_para}
  R(Q^2) =  c_0(W) \, \left[\frac{Q^2}{m_\rho^2} \right]^{c_1}
\end{equation}
with  parameters $c_0=0.32$ and $c_1=0.66$ was obtained.
This parameterisation yields  for the present  exclusive  $\rho^0$ 
data sample an  average value of $D \!=\! 0.4$ at an average 
value of  $R\!=\!0.62$. From the value of $R$ follows that the
contributions of  longitudinal
and transverse photons to $\rho^0$ production  at HERMES 
are about equally important; however, as longitudinal photons 
have zero helicity,   the asymmetry $A_1^\rho$ discussed 
here can arise only from  transverse photons.

In inclusive deep-inelastic scattering,  the asymmetry  $A_2$  was measured
using a transversely polarised target
to be positive but  close to zero~\cite{Adams:1997pr,Abe:1998}.  
In contrast, in  $\rho^0$ production  information about $A_2^\rho$ is available 
through the  measurement of  angular distributions of the decay pions.
For  \mbox{$W\gtrsim 3$~GeV}, the   interference cross section $\sigma_{LT}$ is maximal, since 
the phase difference between amplitudes for  $\rho^0$
production  from transverse and longitudinal photons   was measured to be small~\cite{Hermes:99angles}.  
The contribution of $A_2^\rho$ to $A_\|^\rho$  is  then given 
by the  positivity limit $\!A_2^\rho\! \!= \!\!\sqrt{\!R(Q^2)}$.
It is suppressed by the  small kinematic factor 
$\eta$,  which has an  average value
$\langle \eta \rangle =  0.06$
for the  exclusive $\rho^0$ data sample.

The photon-nucleon asymmetry $A_1^\rho$ in a given kinematic bin  
is  obtained from the experimental lepton-nucleon asymmetry $A_\|^\rho$ using Eq.~\ref{eq:apar2a1}:
\vspace*{-0.05cm}
\begin{equation}  \label{eq:a1etaa2}
  A_1^\rho  =    \frac{A_\|^\rho}{ \langle D \rangle } - \big \langle \eta \sqrt{R(Q^2)} \, \big \rangle,
\end{equation}
\vspace*{-0.05cm}
where  $\langle D \rangle$ and $\big \langle \eta \sqrt{R(Q^2)} \big \rangle$
are the average values for all  events  in the bin.

The stability of the observed asymmetry was studied by
varying the  event selection requirements and 
comparing alternative methods of  background subtraction~\cite{Meissner:phd}.
These systematic studies were   performed 
for both the experimental asymmetry $A_\|^\rho$ and  for 
the  photon-nucleon asymmetry  $A_1^\rho$.
No significant dependence on the 
event selection criteria was observed
during variation of requirements on vertex geometry, particle 
identification and  event kinematics. 
Using the DIS Monte Carlo simulation, a contribution of DIS 
background to  the asymmetry $A^\rho_\|$  of $0.003$ was found, 
and taken as an  upper limit on the associated systematic uncertainty.
Two alternative techniques
of background subtraction  produced similar results: one 
based on events at high  values of $-t'$ where the 
non-exclusive background dominates the data~\cite{Hermes:99coherence},  and
another  using an  empirical fit to the elastic peak 
in the  $\Delta E$ distribution.  
The quoted asymmetries and the bin centers have not been corrected for the 
limited acceptance of the spectrometer for $\rho^0$ production;
resolution effects and  bin-to-bin smearing can be  neglected
due to the large bin size  in the  present analysis.
A comparison of the 1996 and 1997 data sets with opposite beam helicities
yielded consistent results,  excluding  possible 
contributions from single-spin asymmetries.
The contribution of  electroweak radiative processes to
the measured  asymmetry  is expected to  be  negligible~\cite{akus:1999priv}.
The uncertainties in the 
beam and target polarisation measurements cause 
a $6.6\%$ fractional systematic uncertainty in the 
asymmetries. 

Evidence for a positive  double-spin  asymmetry  has been
observed in  the lepton-nucleon  cross section for exclusive $\rho^0$ production  
on the longitudinally polarised hydrogen target.
Averaged over the  kinematic acceptance,
the  measured  asymmetry  has a value of
\begin{equation}
 \langle A_\|^\rho \rangle= 0.119 \pm 0.045_{\rm stat} \pm 0.008_{\rm syst},
\end{equation}
where the  systematic uncertainty is dominated by the contributions
from the  beam and target polarisation measurements. 
Using the average depolarisation factor  $\langle D \rangle \!=\!0.40$,
the average photon-nucleon asymmetry $A_1^\rho$ was found according to Eq.~\ref{eq:a1etaa2}
\vspace*{-0.3cm}
\begin{equation}
  \langle A_1^\rho \rangle = 0.24 \pm 0.11_{\rm stat} \pm 0.02_{\rm syst},
\end{equation}
where  a contribution  $\langle \eta \sqrt{R} \rangle = 0.053$ by the asymmetry $A_2^\rho$
was subtracted.

\begin{figure}[!ht]
 \vspace*{0.1cm}	
  \hspace*{-0.5cm}
  \begin{minipage}{3.8cm}  
    \includegraphics[bbllx=5pt,bblly=15pt,bburx=290pt,bbury=290pt,angle=0,width=4.1cm,height=3.8cm]
    {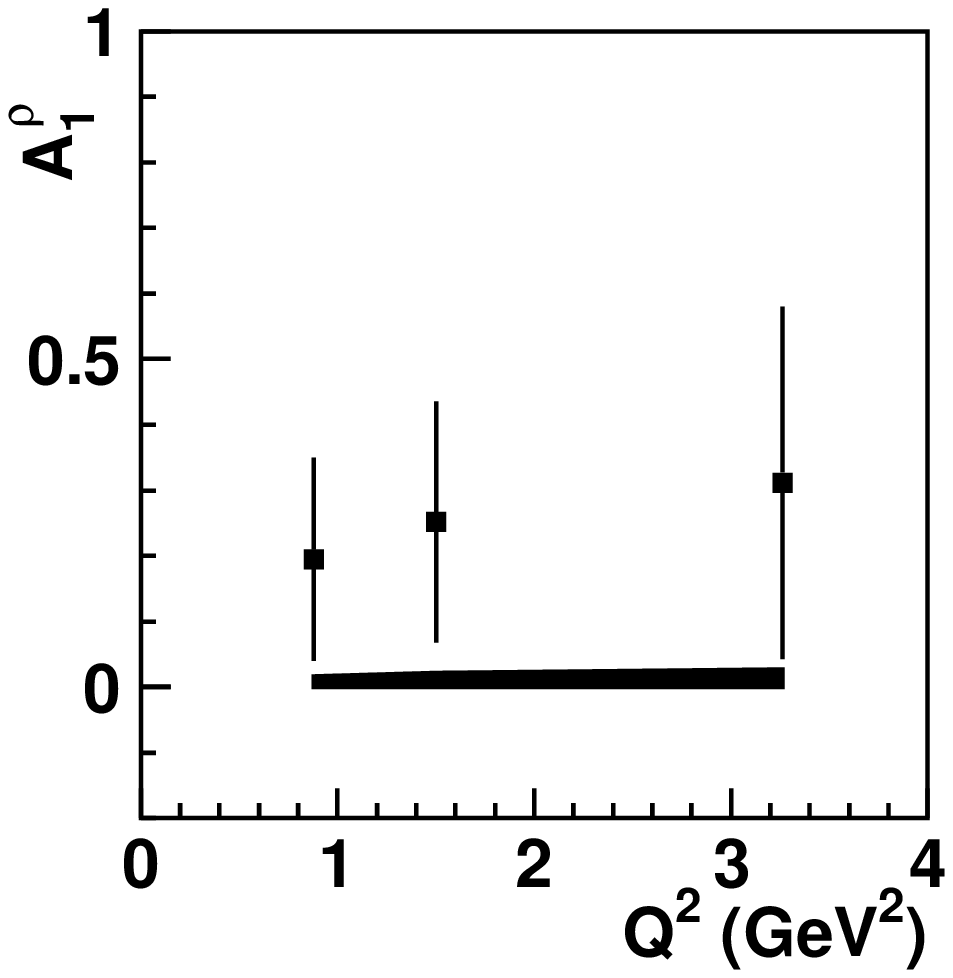}        
  \end{minipage}
  \begin{minipage}{3.8cm} 
    \includegraphics[bbllx=5pt,bblly=15pt,bburx=290pt,bbury=290pt,angle=0,width=4.1cm,height=3.8cm]   {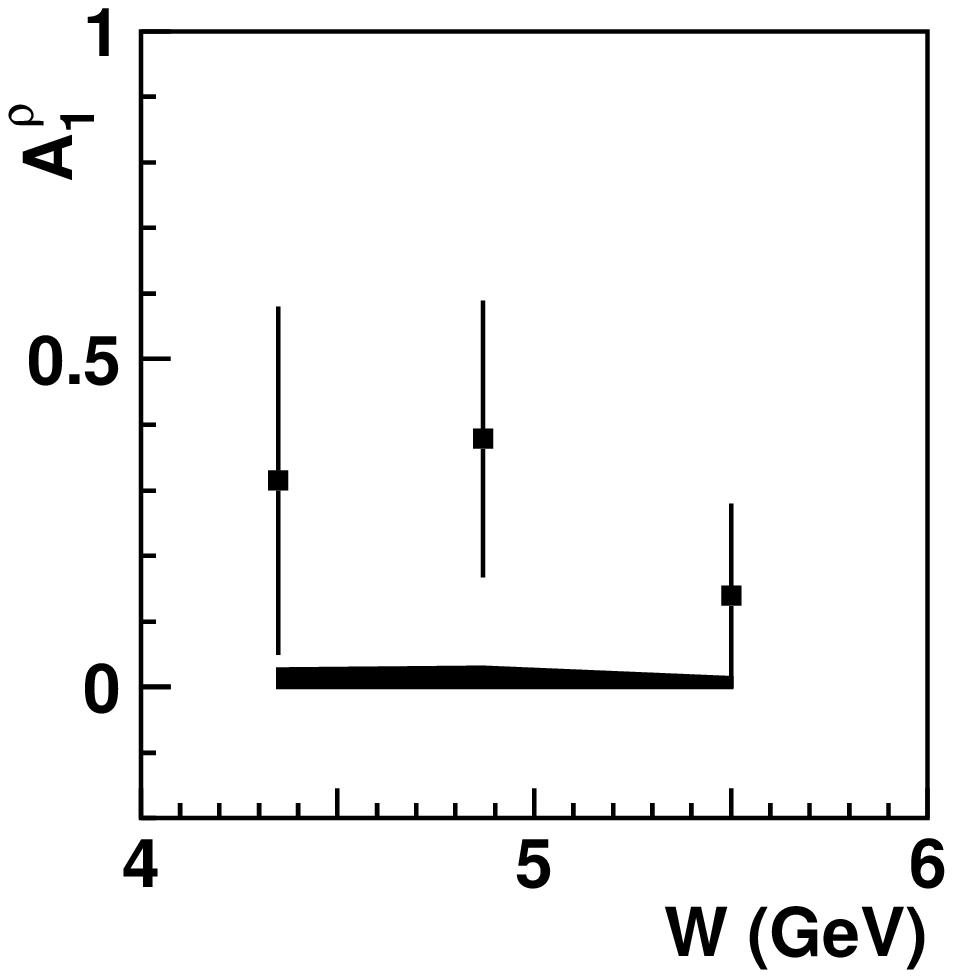} 
  \end{minipage} \\

  \hspace*{-0.5cm}
  \begin{minipage}{3.8cm}  
    \includegraphics[bbllx=5pt,bblly=15pt,bburx=290pt,bbury=290pt,angle=0,width=4.1cm,height=3.8cm]
    {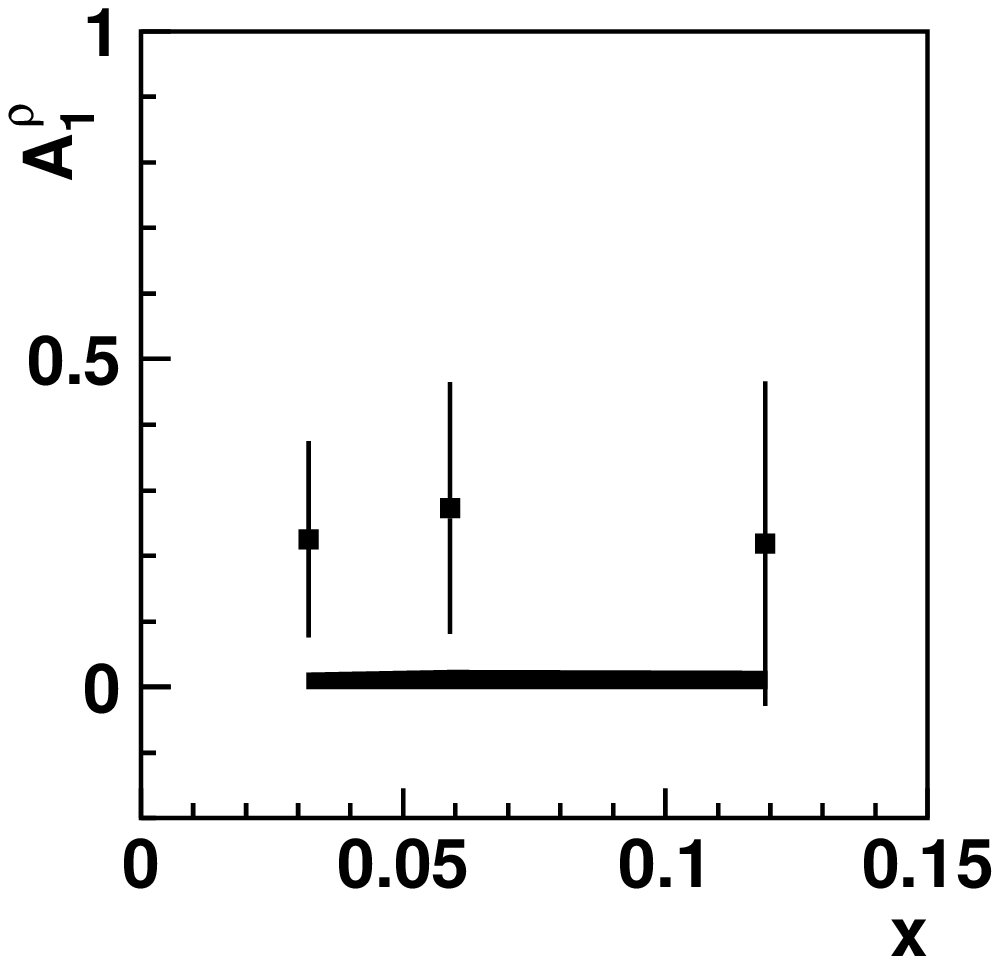}        
  \end{minipage}
  \begin{minipage}{3.8cm} 
    \includegraphics[bbllx=5pt,bblly=15pt,bburx=290pt,bbury=290pt,angle=0,width=4.1cm,height=3.8cm]
    {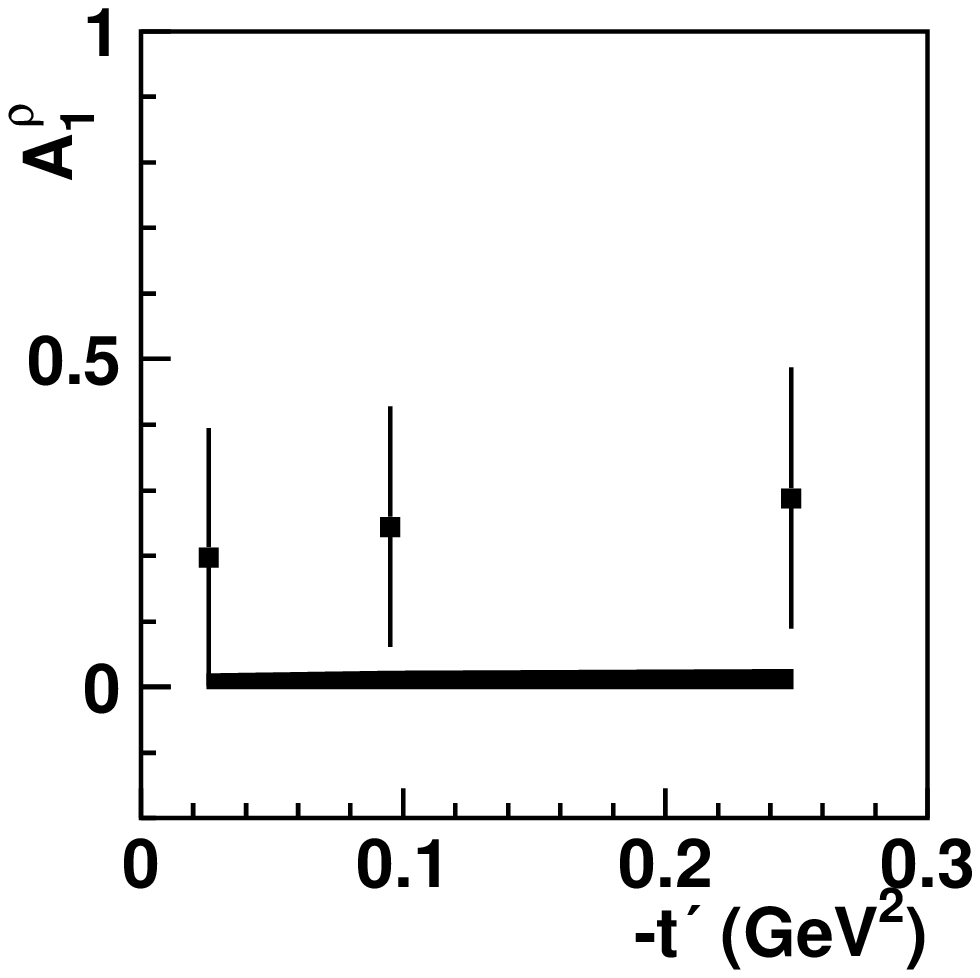} 
  \end{minipage} \\

  \hspace*{-0.5cm}
  \begin{minipage}{3.8cm} 
    \includegraphics[bbllx=5pt,bblly=15pt,bburx=290pt,bbury=290pt,angle=0,width=4.1cm,height=3.8cm]
    {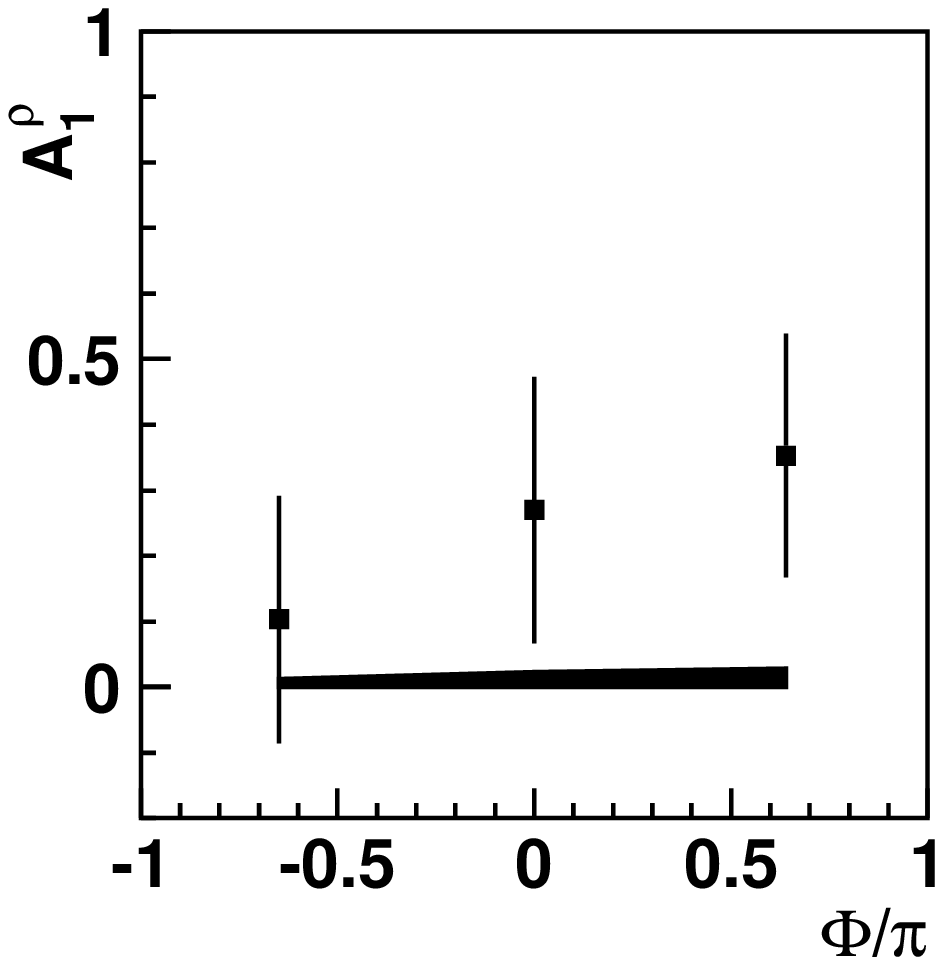}
  \end{minipage}
  \begin{minipage}{3.8cm} 
    \includegraphics[bbllx=5pt,bblly=15pt,bburx=290pt,bbury=290pt,angle=0,width=4.1cm,height=3.8cm]
    {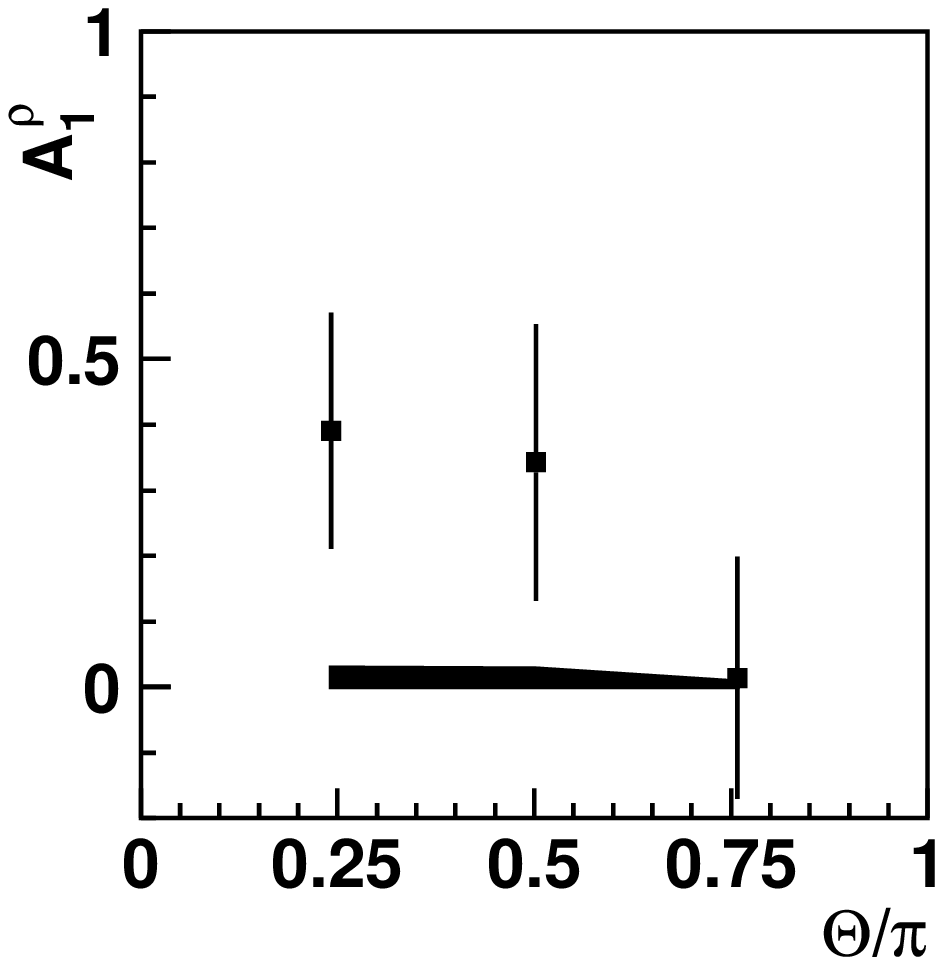}
  \end{minipage} \\
  \caption{ \label{fig:a1etaa2} 
    Photon-nucleon  asymmetry  $A_1^\rho$ in exclusive $\rho^0$ production 
    versus $Q^2$, $W$
        $x$, $-t'$, $\Phi$, and $\theta$.
    Error bars and error bands denote the statistical
    and experimental systematic uncertainties, respectively. }
 \end{figure}

Fig.~\ref{fig:a1etaa2} shows the dependence of   $A_1^\rho$
on the kinematic variables  $Q^2$, $W$, $x$, $-t'$,
and the angles $\Phi$ and $\theta$.
Error bars denote the statistical uncertainties and the dark  band
at the bottom of the plots indicates the   systematic
uncertainty.
Within the statistical uncertainty of the measurement,
no significant dependence on  any of the kinematic 
variables or angles   can be seen in either  $A_\|^\rho$ or in $A_1^\rho$. \\

The present result  for $A_1^\rho$ is compared to 
the already mentioned theoretical prediction~\cite{Fraas:1976asym},
which is based on the description of diffractive 
exclusive $\rho^0$ leptoproduction 
{\it and} inclusive deep-inelastic  \linebreak
scattering  by the GVMD.
At $x<0.2$,  this model can  relate the asymmetry for 
exclusive $\rho^0$ production  $A_1^\rho$
to the asymmetry $A_1^{\gamma^\star p}$   for inclusive  
DIS  at the same value of $x$. 
Assuming an  {\it approximate} validity of SCHC,  spin asymmetries were 
written as the ratio  of he\-licity-conserving  amplitudes of unnatural $(-1)^{L+1}$ to
natural $(-1)^{L}$ parity exchange  in the $t$-channel.
A non-zero asymmetry indicates a  contribution  
of exchange processes with unnatural  parity to the {\it interference} 
responsible for the asymmetry. 
This contribution may be large enough to yield 
an asymmetry while remaining negligible  in the
{\it incoherent sum} of squared amplitudes for the cross section
that was observed  to be dominated by natural parity exchange
in measurements of angular distributions 
in $\rho^0$ production and decay~\cite{Hermes:99angles}. 
Such an unnatural parity exchange is consistent with  an exchange of diquark objects 
as they can have both natural and unnatural parity.
Quark exchange has been shown to be the predominant contribution
in exclusive  $\rho^0$ production by longitudinal photons at HERMES 
energies~\cite{Hermes:00rhoxsect}; the present result suggests 
that diquark exchange may also contribute to $\rho^0$ production
from transverse photons.

In Ref.~\cite{Fraas:1976asym}, numerical predictions for the ratio 
$A_1^\rho/A_1^{\gamma^\star p}$ were made 
for lepton beam energies of $15$~GeV and $50$~GeV. 
From these predictions,  a ratio  $A_1^\rho/A_1^{\gamma^\star p}$
of about $2$  can be interpolated for the  HERMES beam energy of 27.56~GeV i.e.
at $\langle W \rangle =4.9$~GeV.
With a value of $A_1^{\gamma^\star p}(x\!=\!0.07)\!=\!0.13$,  obtained from a 
parameterisation  of  the inclusive asymmetry measured at HERMES~\cite{Hermes:98g1p},
a ratio  $A_1^\rho/A_1^{\gamma^\star p}\!= \!1.9 \!\pm\! 0.8$ is inferred from 
the data, where  
the  uncertainty is  determined  from  the statistical uncertainty of $A_1^\rho$.
This result  is consistent with the above 
theoretical  prediction that was made a quarter of a century 
before the data became available.

\begin{figure}[!ht]
  \includegraphics[width=7.5cm,height=2.8cm]{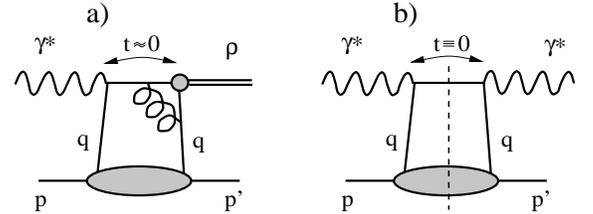}        
  \caption{Schematic graphs for exclusive $\rho^0$  production (a) and 
inclusive lepton-nucleon scattering (b).
    \label{fig:ArhoAdis}}
\end{figure}

The  ratio  $A_1^\rho/A_1^{\gamma^\star p}$  also has another 
interpretation if certain assumptions are made.
The two  processes are schematically shown in Fig.~\ref{fig:ArhoAdis}.
As already mentioned, only transverse photons can contribute 
to the double-spin asymmetries $A_1^\rho$ and 
$A_1^{\gamma^\star p}$.
For transverse photons both asymmetries can be expressed in terms of $s$-channel 
helicity amplitudes as
\begin{equation} 
 \hspace*{-3mm}
 A_1^{\rho}\!\!=\! \frac{|f_{++}^{++}|^2 \!\!-\! |f_{++}^{--}|^2}{ |f_{++}^{++}|^2 \!\!+\! |f_{++}^{--}|^2},
\;
  A_1^{\gamma^*\!p} \!\!=\! \frac{{\rm Im}f_{++}^{++} \!\!-\! {\rm Im}f_{++}^{--}} {{\rm Im}f_{++}^{++} \!\!+\! {\rm Im}f_{++}^{--}}.
\hspace*{-3mm}
\end{equation}
Here,  $f^{\alpha \alpha'}_{i i'}$ 
are the helicity-conserving ($\alpha \!=\!\alpha'$, $i\!=\!i'$) amplitudes 
with  $i,i'=\pm 1/2$ the helicity of the incident and scattered nucleon, $\alpha \!=\! 0,\pm 1$
the helicity of the incident photon, and $\alpha '\!=\! 0,\pm 1$ the helicity of the 
outgoing photon or $\rho^0$ meson.
Here it is assumed  that  the above helicity  amplitudes
for $\rho^0$ production and inclusive reactions  differ only by a common factor, 
and that  the contributions of helicity-flip amplitudes 
($\alpha \!\ne\! \alpha'$,  $i \! \ne\! i'$) 
are small and can be  neglected.
For the asymmetry in the cross section for $\rho^0$ production,
the amplitudes have to be squared, while the expression for the  asymmetry in
the inclusive reaction is  based on the optical theorem.
Assuming  that  $f_{++}^{++}$ and  $f_{++}^{--}$ are primarily imaginary, as is the 
case for exclusive reactions at small $x$,
the ratio $A_1^\rho/A_1^{\gamma^*p}$  can be approximated as 
\vspace*{-0.1mm}
\begin{equation} \label{eq:ArhoA1} 
          \frac{A_1^{\rho}}{A_1^{\gamma^*p }} \simeq \frac{2}{1+(A_1^{\gamma^*p })^2},
\end{equation}
which is about $2$ since the inclusive asymmetry at low $x$ is of order $0.1$.

In models based on perturbative QCD~\cite{Collins:1997},
dif\-fractive $\rho^0$ production  by longitudinal photons is described by 
three distinct components:  a distribution amplitude  for the  meson, 
a hard scattering  amplitude for the exchange of quarks and gluons,
and a  non-perturbative  description of the target nucleon by 
skewed parton distributions (SPD's) allowing a sensitivity of the diffractive process 
to the internal (spin)  structure of the nucleon.
A general proof of the  factorisation theorem in diffractive 
meson production, an important  prerequisite  for 
the definition of SPD's,
exists  only for {\it longitudinal} photons,
and does not   apply to  the production of 
mesons from {\it transverse} photons~\cite{Collins:1997,Mankiewicz:1999-1}.
As an important consequence,  a clear interpretation of the asymmetry  $A_1^\rho$ 
within the framework of perturbative QCD   and skewed parton distributions 
presently does not exist and would require substantial theoretical progress. 

The present result  indicating  a non-zero  double-spin  asymmetry 
is in contrast  to  the  preliminary result  
of a similar measurement by the  SMC collaboration~\cite{SMC:rhoA1-DIS99} at 
comparable values of $Q^2$ but at three times  higher $W$, i.e. at smaller $x$.
Their measurement  of $A_\|^\rho$   in  several bins in $Q^2$ is consistent with   zero,
with  a better  statistical precision  than the present measurement. 
In the context of Ref.~\cite{Fraas:1976asym} their $A_\|^\rho$ is expected to be smaller  since 
the asymmetries in inclusive deep-inelastic scattering decrease at the  smaller values of $x$
probed by the SMC measurement.
Alternatively, at  SMC the diffractive process  
is believed to be dominated by Pomeron or gluon exchange,
whereas at the lower HERMES energy, there are indications that 
$\rho^0$ production is dominated by Reggeon
or quark exchange~\cite{vanderh,Hermes:00rhoxsect}; 
the different   asymmetry results of SMC and HERMES  might well reflect the
different production  mechanisms in the two kinematic regimes.

In summary, evidence for  a non-zero longitudinal double-spin
asymmetry in the cross section for  exclusive  $\rho^0$  
production in  lepton-nucleon scattering has been observed.
The value measured  is $\langle A_1^\rho  \rangle = 
0.24 \pm 0.11_{\rm stat} \pm 0.02_{\rm syst}$,
at 
$\langle W \rangle = 4.9$~GeV  and 
$\langle Q^2 \rangle = 1.7$~GeV$^2$.  
No  significant dependence  
on any kinematic variable was observed.
A ratio  $A_1^\rho/A_1^{\gamma^\star p}= 1.9 \pm 0.8$
of the asymmetry  in $\rho^0$  production to that in inclusive lepton-nucleon scattering was obtained.
This  result is consistent with an early  prediction based 
on the generalised vector meson dominance model. \\

We gratefully acknowledge the DESY management for its support and
the DESY staff and the staffs of the collaborating institutions.
This work was supported by the FWO-Flanders, Belgium;
the Natural Sciences and Engineering Research Council of Canada;
the INTAS and TMR network contributions from the European Community;
the German Bundesministerium f\"ur Bildung und For\-schung;
the Deutsche Forschungsgemeinschaft (DFG); 
the Deutscher Akademischer Austauschdienst(DAAD);
the Italian Istituto Nazio\-nale di Fisica Nucleare (INFN);
Monbusho International Scientific  Research Program, JSPS, and Toray
Science Foundation of Japan;
the Dutch Foundation for Fundamenteel Onderzoek der Materie (FOM);
the U.K. Particle Physics and Astronomy Research Council; and
the U.S. Department of Energy and National Science Foundation.


\begin{thebibliography}{}
\bibitem{Sakurai} J.J.~Sakurai, Ann. Phys. {\bf 11} (1960) 1, 
and  Phys. Rev. Lett. {\bf 22} (1969) 981.
\bibitem{Bauer:1978} T.H.~Bauer,  R.D.~Spital, D.R.~Yennie and F.M.~Pipkin, Rev. Mod. Phys. {\bf 50} (1978) 261.
\bibitem{Cassel:1981} D.G.~Cassel {\it et al.}, Phys. Rev. {\bf D24} (1981) 2787.
\bibitem{Haakman:1996} L.P.A.~Haakman, A. Kaidalov and J.H.Koch, Phys. Lett. {\bf B365} (1996) 411.
\bibitem{Crittenden:1997} J.A.~Crittenden, Springer Tracts in Modern Physics 140 (1997).
\bibitem{Hermes:00rhoxsect} HERMES Collaboration, A.~Airapetian {\it et al.}, Eur. Phys. J. {\bf C17} (2000) 389.
\bibitem{rady:1996} A.V.~Radyushkin, Phys. Lett. {\bf B380} (1996) 417, 
                        and Phys. Rev. {\bf D56} (1996) 5524.
\bibitem{XJi:1997} X.~Ji, Phys. Rev.   {\bf D55} (1997) 7114, 
                        and Phys. Rev. Lett. {\bf 78} (1997) 610.    
\bibitem{Collins:1997}  J.C.~Collins, L.~Frankfurt and  M.~Strikman, Phys. Rev. {\bf D56} (1997) 2982.    
\bibitem{vanderh}  M.~Vanderhaeghen, P.A.M.~Guichon and M.~Guidal, Phys. Rev. Lett. {\bf 80} (1998) 5064, and    Phys. Rev. {\bf D60} (1999) 094017.    
\bibitem{Zeus:rho} ZEUS Collaboration, J.Breitweg {\it et al.},  Eur. Phys. J.  {\bf C2} (1998) 247.
\bibitem{H1:rho} H1 Collaboration, C.~Adloff {\it et al.}, Eur. Phys. J. {\bf C13} (2000) 371.
\bibitem{wolf:1973}  K.~Schilling and  G.~Wolf, Nucl. Phys. {\bf B61} (1973) 381.    
\bibitem{Fraas:1974} H.~Fraas, Ann. Phys. {\bf 87} (1974) 417.
\bibitem{Fraas:1975gvmd} H.~Fraas, B.J.~Read and D.~Schildknecht, Nucl. Phys. {\bf B86} (1975) 346, 
                        and Nucl. Phys. {\bf B88} (1975) 301.
\bibitem{Fraas:1976asym} H.~Fraas, Nucl. Phys. {\bf B113} (1976) 532.
\bibitem{vaenttinen:1998} M.~V{\"a}nttinen and L.~Mankiewicz, Phys. Lett. {\bf B434} (1998) 141, and Phys. Lett. {\bf B440} (1998) 157.
\bibitem{Hermes:98spec} HERMES Collaboration, K.~Ackerstaff  {\it et al.}, Nucl. Inst. Meth. {\bf A417} (1998) 230.
\bibitem{Sokolov-1964}  A.A.~Sokolov and I.M.~Ternov,  Sov. Phys. Doklady {\bf 8} (1964) 1203.


\bibitem{bar}  D.P.~Barber {\em et al.}, Nucl. Instr. and Meth. {\bf A 329} (1993)
79; A. Most, Proc. of the 12th International Symposium on High-Energy Spin Physics,
edited by C.W. de Jager {\em et al.},  Amsterdam, The Netherlands,
World Scientific (1997) 800.

\bibitem{ste} J.~Stewart, Proc. of the Workshop Polarised gas targets and
polarised beams, edited by R.J.~Holt and M.A.~Miller, Urbana-Champaign, USA,
AIP Conf. Proc. {\bf 421} (1997) 69.

\bibitem{Hermes:99coherence} HERMES Collaboration, K.~Ackerstaff {\it et al.}, Phys. Rev. Lett. {\bf 82} (1999) 3025.

\bibitem{LEPTO} B.~Anderson, G.~Gustafson, G.~Ingelman and T.~Sjoestrand, Z. Phys. {\bf C9} (1981) 233.    
\bibitem{LUND:94} T.~Sj{\"o}strand, Comp. Phys. Comm. {\bf 82} (1994)  74.
\bibitem{E665:1997} E665 Collaboration, M.R.~Adams {\it et al.}, Z. Phys. {\bf C74} (1997) 237.
\bibitem{Withlow:90} L.W.~Whitlow {\it et al.}, Phys. Lett. {\bf B250} (1990) 193.
\bibitem{Hermes:99angles}  HERMES Collaboration, 
K.~Ackerstaff  {\it et al}, Eur. Phys. Jour. C (in press), hep-ex/0002016. 
\bibitem{Abe:1998} E143 Collaboration, K.~Abe {\it et al.}, Phys. Rev.  {\bf D58} (1998) 112003.
\bibitem{Adams:1997pr} SMC Collaboration, D.~Adams {\it et al.}, Phys. Rev. {\bf D56} (1997) 5330.
\bibitem{Meissner:phd} F. Meissner, PhD Thesis, Humboldt University Berlin (2000), DESY-THESIS 2000-014. 
\bibitem{akus:1999priv} I.~Akushevich, Private Communication (1999).
\bibitem{Hermes:98g1p} HERMES Collaboration, A.~Airapetian {\it et al.}, Phys. Lett. {\bf B442} (1998) 484.
\bibitem{SMC:rhoA1-DIS99} A.~Tripet, Nucl. Phys. {\bf B (Proc. Suppl.) 79} (1999) 529. 
\bibitem{Mankiewicz:1999-1} L.~Mankiewicz and G.~Piller, Phys. Rev. {\bf D61} (1999) 074013.
\bibitem{Brodsky:1994} S.J.~Brodsky {\it et al.},
        Phys. Rev. {\bf D50} (1994)  3134.    
\end{thebibliography}
\end{document}